# Observational study of the formation of homologous confined circular-ribbon flares

Shuhong Yang[1,2], Ruisheng Zheng[3], Yijun Hou[1,2], Yuandeng Shen[4,5], Yin Li[1,2], Xiaoshuai Zhu[6], Ting Li[1,2], and Guiping Zhou[1,2]

[1] State Key Laboratory of Solar Activity and Space Weather, National Astronomical Observatories, Chinese Academy of Sciences, Beijing 100101, People's Republic of China; e-mail: shuhongyang@nao.cas.cn

[2] School of Astronomy and Space Science, University of Chinese Academy of Sciences, Beijing 100049, People's Republic of China

[3] Institute of Space Sciences, Shandong University, Weihai 264209, People's Republic of China

[4] State Key Laboratory of Solar Activity and Space Weather, School of Aerospace, Harbin Institute of Technology, Shenzhen 518055, People's Republic of China

[5] Shenzhen Key Laboratory of Numerical Prediction for Space Storm, Harbin Institute of Technology, Shenzhen 518055, People's Republic of China

[6] State Key Laboratory of Solar Activity and Space Weather, National Space Science Center, Chinese Academy of Sciences, Beijing 100190, People's Republic of China

**ABSTRACT**

Context. When several solar flares with comparable classes occur successively at the same location and exhibit similar morphological features, they are called homologous flares. During 2012 May 8-10, five M-class homologous circular-ribbon flares associated with no coronal mass ejection occurred in active region (AR) 11476. The formation process of these homologous confined flares, particularly the homologous aspect, is unclear and inconclusive.
Aims. This paper is dedicated to studying how the energy for this series of flares was accumulated and whether there existed null points responsible for the flare energy release.
Methods. With the multi-wavelength images and vector magnetograms from the Solar Dynamics Observatory, we study the formation process of these homologous confined circular-ribbon flares. Using the nonlinear force-free field modeling, the three-dimensional coronal magnetic structures are reconstructed.
Results. Before and during the five flares, the sunspots with opposite polarities sheared against each other and also rotated individually. Before each flare, the magnetic fields at the polarity inversion line were highly sheared and there existed a magnetic flux rope overlain by arch-shaped loops. For the first four flares, we find magnetic null points in the fan-spine topology, situated at about 3.8 Mm, 5.7 Mm, 3.4 Mm, and 2.6 Mm above the photosphere, respectively. For the fifth flare, no null point is detected. However, in the (extreme-)ultraviolet images, the evolution behaviors of all the flares were almost identical. Therefore, we speculate that a null point responsible for the occurrence of the fifth flare may have existed.
Conclusions. These results reveal that, for these homologous flares in AR 11476, the sunspot rotation and shearing motion play important roles in energy accumulation, the null point of the fan-spine topology is crucial for energy release through magnetic reconnection therein, and large-scale magnetic loops prevent the erupting material from escaping the Sun, thus forming the observed homologous confined major circular-ribbon flares. This study provides clear evidence for the drivers of successive, homologous flares as well as the nature of confined events.

## 1. Introduction

Solar flares are one kind of energetic phenomenon on the Sun and have been extensively investigated (e.g., Priest & Forbes 2002; Yang et al. 2017). During solar flares, magnetic energy is suddenly released through the process of magnetic reconnection (Masuda et al. 1994; Su et al. 2013). Magnetic reconnection results in the rearrangement of magnetic topology and converts the free energy of magnetic field into the kinetic and thermal energy of plasma (Zweibel & Yamada 2009; Yang et al. 2015; Yang & Xiang 2016). Solar flares associated with coronal mass ejections (CMEs) are termed eruptive flares, while those without CMEs are called confined flares (Schrijver 2009; Yang et al. 2014b, 2019; Li et al. 2020; Kazachenko 2023). Understanding the properties of confined flares is vital for improving our ability to predict solar activity and space weather (Temmer 2021).

Magnetic free energy is stored in non-potential magnetic fields, and solar flares tend to take place at sites with high non-potentiality (Schrijver et al. 2008). Before flares, the magnetic fields around the polarity inversion line (PIL), i.e., the separation line between magnetic fields with opposite polarities, are characterized by strong gradients and large shear angles (Toriumi & Wang 2019). Rapid shearing flows in the photosphere play an important role in accumulating the free energy that can be released to power solar flares (Meunier & Kosovichev 2003; Shimizu et al. 2014; Park et al. 2018; Chintzoglou et al. 2019). Sunspot rotation is also found to be crucial for the storage and release of magnetic free energy. Due to the rotation of sunspots, the magnetic field lines that connect the positive and negative sunspots are twisted, sometimes even forming flux ropes (Masson et al. 2009; Yan et al. 2009; Inoue et al. 2012; Amari et al. 2014; Gopasyuk 2015; Zhang et al. 2022). Vemareddy et al. (2016) found that the time evolution of non-potential magnetic parameters closely corresponds to the rotation of sunspots.

As one type of solar flare, circular-ribbon flares are related with the presence of fan-spine topology (Sun et al. 2013; Wyper et al. 2017; Li et al. 2018; Hou et al. 2019; Yang et al. 2020b; Mitra et al. 2023). In the fan-spine magnetic configuration, two distinct connectivity domains are separated by a dome-shaped fan with the outer and inner spines converging at the null point (Shibata et al. 1994; Pontin et al. 2013; Yang et al. 2020b). The fan-spine magnetic topology is favorable for the occurrence of solar flares via magnetic reconnection at the null point (Priest & Titov 1996; Antiochos 1998; Aulanier et al. 2000; Cheng et al. 2023). When magnetic reconnection occurs at the null point of the fan-spine, high-speed particles move down along the fan and then hit the lower atmosphere, forming a circular-shaped ribbon (Masson et al. 2009, 2017; Reid et al. 2012; Shen et al. 2019; Lee et al. 2020; Zhang et al. 2020; Zhang 2024).

In the case of circular-ribbon flares, the magnetic flux rope embedded within the fan dome is considered to play a crucial role in producing flares (Joshi et al. 2015; Liu et al. 2015). If a magnetic flux rope is filled with some dense and low-temperature material, it appears as a filament (Parenti 2014; Yang et al. 2014a; Sterling et al. 2015; Zheng et al. 2019). When the flux rope starts to ascend, a breakout-type reconnection at the null point of the fan-spine structure occurs. During this process, the loops overlying the flux rope are stretched to create a current sheet, where magnetic reconnection also occurs (Shibata et al. 1995; Lin & Forbes 2000). Thus a pair of parallel ribbons is formed within a0 circular ribbon (Hernandez-Perez et al. 2017). If the spine is connected to a remote solar surface, then the material ejected from the reconnection site flows along the closed field lines and eventually reaches the far end causing a remote brightening (Wang & Liu 2012; Deng et al. 2013; Yang & Zhang 2018; Yang et al. 2020b).

When several solar flares with comparable classes occur successively at the same location and exhibit similar morphological features, they are called homologous flares (Martres 1989; Choe & Cheng 2000; Chandra et al. 2011). During 2012 May 8-10, a series of homologous confined major circular-ribbon flares occurred in active region (AR) 11476. Then one may want to know how the energy for this series of flares was accumulated and whether there existed null points responsible for the flare energy release. Therefore, this paper is dedicated to studying the formation process of these flares, which is more related to the homologous aspect rather than the confined aspect.

## 2. Observations

The Helioseismic and Magnetic Imager (HMI; Scherrer et al. 2012; Schou et al. 2012) on board the Solar Dynamics Observatory (SDO; Pesnell et al. 2012) measures the full-disk magnetic fields with a pixel size of $0''.5$. We use five series of the line-of-sight (LOS) magnetograms and the intensitygrams with a cadence of 45 seconds around the peak times of the five flares. The periods of the data sets are 13:00-13:15 UT on May 8, 12:20-12:37 UT on May 9, 21:00-21:10 UT on May 9, 04:10-04:25 UT on May 10, and 20:20-20:30 UT on May 10. The SDO Atmospheric Imaging Assembly (AIA; Lemen et al. 2012) provides the full-disk images at ten different wavelengths with a spatial sampling of $0''.6$ pixel$^{-1}$. The AIA ultraviolet (UV) and extreme-ultraviolet (EUV) images have cadences of 24 seconds and 12 seconds, respectively. We adopt 1600 Å, 304 Å, 171 Å, and 94 Å images. The five sets of HMI and AIA data are calibrated to Level 1.5 using the standard routine within the Solar Software package, and then each set of data is differentially rotated to the reference time corresponding to the start of the respective flare. In addition, we employ the 1-min cadence data from the Geostationary Operational Environmental Satellite (GOES) to examine the variation of soft X-ray 1–8 Å flux.

Moreover, we use vector magnetograms and corresponding continuum maps of the Space Weather HMI AR Patches (SHARP; Bobra et al. 2014) from 12:00 UT on May 7 to 23:48 May 10 with a cadence of 12 min. With the SHARP vector magnetograms as the bottom boundary, we reconstruct the coronal magnetic structures using the nonlinear force-free field (NLFFF) modeling (Wheatland et al. 2000; Wiegelmann et al. 2012). Before the start of each flare, an observed vector magnetogram is adopted and preprocessed in order to best suit the force-free conditions (Wiegelmann et al. 2006). With the help of an optimization method, the preprocessing procedure modifies the transverse magnetic field components within the noise level to eliminate the net Lorentz force and torque in the original vector magnetogram. This generates a boundary condition that satisfies the force-free assumption required NLFFF extrapolation while maintaining minimal deviation from the original measurements. NLFFF extrapolations provide critical insights into active regions by reconstructing magnetic configurations under the force-free assumption. While the static nature of NLFFF models limits their applicability during the impulsive flare phase, where dynamic effects dominate, they remain effective for diagnosing pre-flare energy storage and post-flare structural relaxation. Thus, NLFFF extrapolation serves as a foundational tool for probing quasi-static equilibrium states flanking flare activity. We perform the NLFFF extrapolation in the cubic box of $352 \times 224 \times 200$ uniformly spaced grid points with $\Delta x = \Delta y = \Delta z = 1''$. For the extrapolated coronal magnetic fields, we calculate the twist number $\mathcal{T}$ (Berger & Prior 2006) and the squashing factor $Q$ (Demoulin et al. 1996; Titov et al. 2002) using the code developed by Liu et al. (2016).

## 3. Results

The overview of AR 11476 where the five major flares took place is shown in Figure 1. HMI observations show that the polarities of the leading and trailing sunspots (panel (a)) of AR 11476 are negative and positive (panel (b)), respectively. The five M-class flares we study occurred at the location of a bipole (denoted by the arrows in the top panels) within the trailing positive fields. The magnetic strength of this bipolar region exceeds 1 kG, as revealed in the HMI magnetogram. This strength is strong enough to form a pair of sunspots, which are visible in the HMI intensitygram. In the AIA 304 Å image (panel (c)), there is a conspicuous filament (indicated by the green arrow) above the PIL of the bipole. While in the AIA 171 Å image (panel (d)), in addition to the existence of the dark filament, the leading and trailing magnetic polarities of AR 11476 are connected by a set of large-scale coronal loops (denoted by the white arrow).

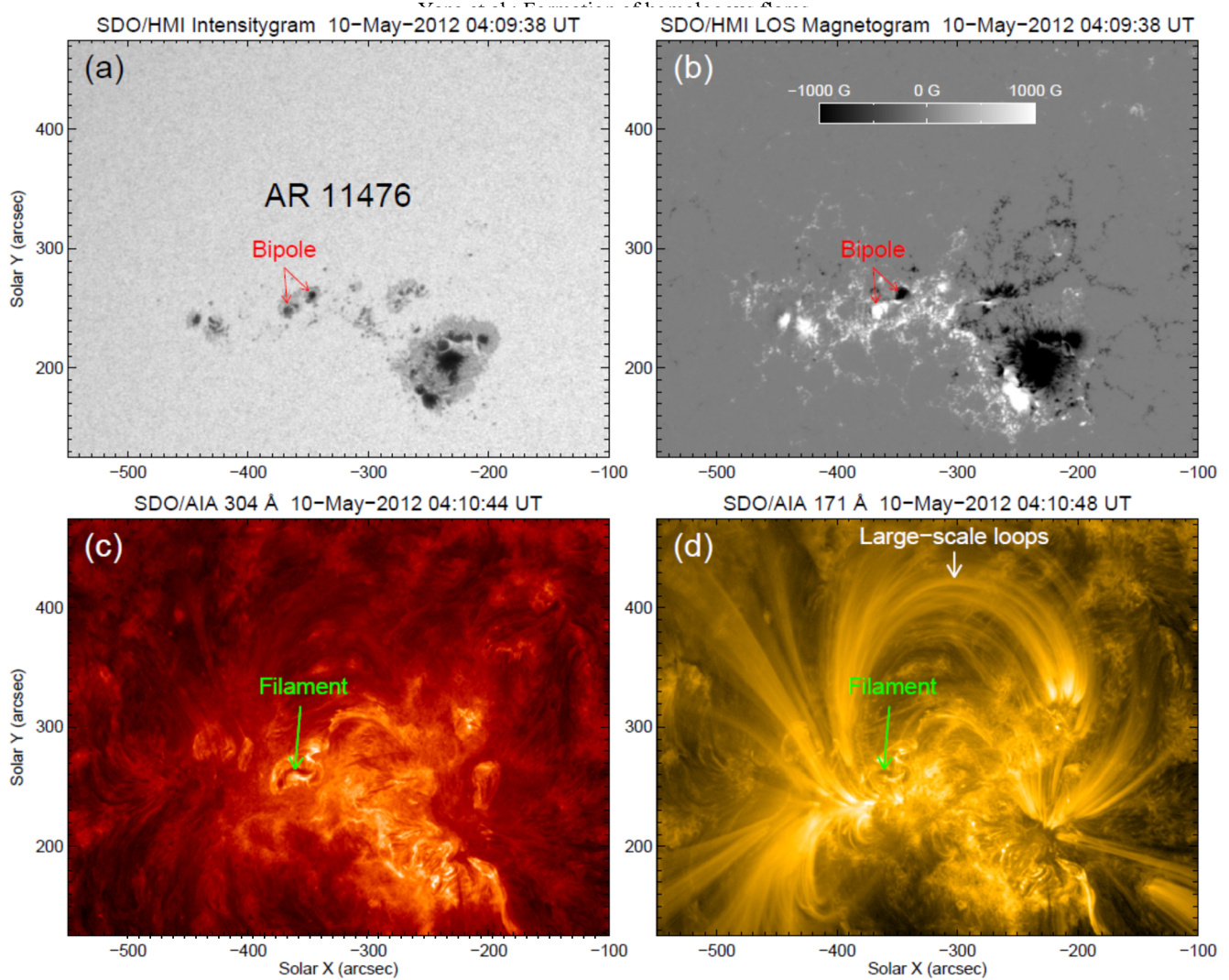

**Fig. 1.** Overview of AR 11476 on 2012 May 10. (a-b) HMI intensitygram and LOS magnetogram displaying the sunspots and the corresponding magnetic fields in the photosphere. (c-d) AIA 304 Å and 171 Å images showing the appearance of overlying structures. The arrows in (a-b) denote a pair of sunspots with opposite polarities, i.e., a bipole. The green arrows in (c-d) indicate a filament, and the white arrow in (d) denotes a set of large-scale loops.

As revealed by the GOES soft X-ray flux in the range of 1–8 Å (shown in the top panels of Figure 2), the five M-class flares are M1.4 (hereafter Flare 1), M4.7 (Flare 2), M4.1 (Flare 3), M5.7 (Flare 4), and M1.7 (Flare 5), and the corresponding durations are 10 min, 15 min, 8 min, 12 min, and 10 min, respectively (see Table 1). These five flares took place at the same location with similar morphologies (see panels (b1)-(b5)). For each flare, there was an erupting filament (denoted by arrow "1"), a circular flare ribbon (indicated by arrow "2"), a pair of parallel flare ribbons (indicated by arrows "3"), and a remote brightening (denoted by arrow "4") as revealed in the AIA 1600 Å images. While in the AIA 94 Å images (the bottom panels), each flare featured a bright kernel (indicated by the red arrow), which corresponds to the post-flare loops connecting the two parallel flare ribbons. Additionally, a cluster of large-scale loops (denoted by the white arrow) was observed overlying the erupting filament and extended to the remote brightening area. Since all the five major flares occurred successively at the same location, have comparable classes (M1.4, M4.7, M4.1, M5.7, and M1.7), exhibit similar morphologies, and were not associated with any CME, they can be termed homologous confined major circular-ribbon flares.

In order to study the energy buildup of the flares, we examine the sunspot evolution in the photosphere (see Figure 3 and Movie 1). In each HMI intensitygram, the green and red plus symbols represent the intensity centroids of sunspots with the positive (P) and negative (N) polarities, respectively. At 11:58 UT on 2012 May 7, the orientation from "P" to "N" was 247° (panel (a)). The sunspots with different polarities sheared in a clockwise direction, and 25 hr later, the orientation changed to 169° (panel (b)). The averaged shearing velocity was 3.1° hr$^{-1}$. Then the clockwise shearing continued, and approximately 23 hr later, the orientation became 55° (panel (c)). The shearing velocity increased to 4.9° hr$^{-1}$. Subsequently, the shearing motion began slow down to 2.2° hr$^{-1}$ and the distance between the two centroids increased significantly (panel (d)). In the following day, the shearing velocity was only about 0.43° hr$^{-1}$.

Besides the shearing motion between the sunspots with opposite polarities, the positive and negative sunspots also underwent individual rotations (see also Movie 1). With the differential affine velocity estimator method (Schuck 2006), we calculate the horizontal velocities of the photospheric structures in the HMI intensitygrams (Figure 4). The horizontal velocity field shown in panel (a) is estimated with two intensitygrams observed at 18:22 UT and 18:34 UT on May 9, and that in panel (b) with two intensitygrams observed at 03:22 UT and 03:34 UT on May 10. The arrows represent the moving velocities of different parts of the main bipolar patches. As revealed by the arrow directions, the positive polarity sunspot (blue contour at 500 G) showed persistent rotation, appearing as a pronounced counterclockwise vortex structure. Although the negative polarity sunspot (red contour at -500 G) displayed slight clockwise rotation during some certain periods (panel (a)), it predominantly rotated counterclockwise, manifesting as a distinct vortex (panel (b)). As we can see in Movie 1, the torsional motion occurred prior to all flares.

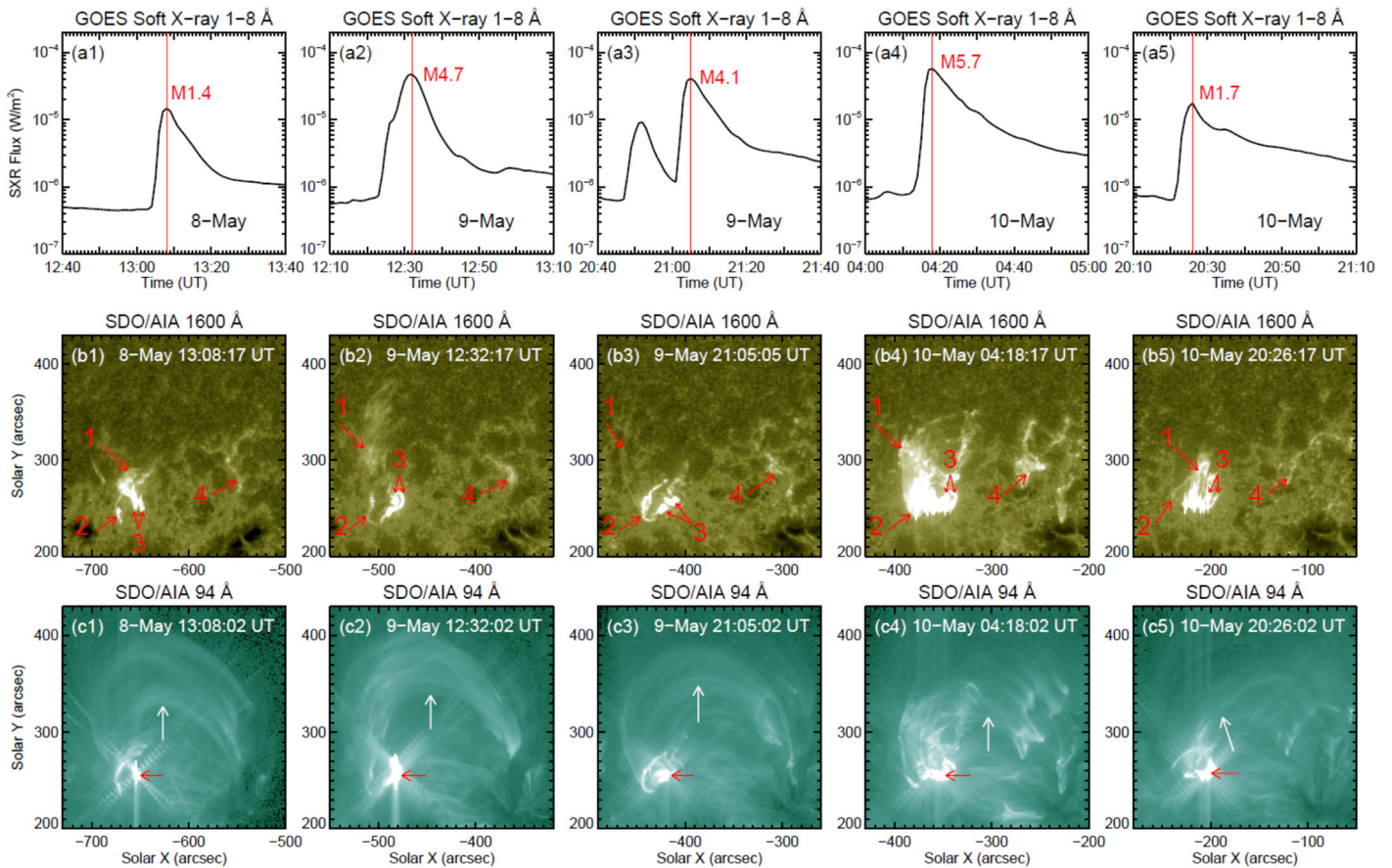

**Fig. 2.** Appearance of the five homologous flares. (a1-a5) GOES soft X-ray flux in 1–8 Å showing the M1.4, M4.7, M4.1, M5.7, and M1.7 flares. (b1-b5) AIA 1600 Å and (c1-c5) AIA 94 Å images showing the five flares in UV and EUV wavelengths, respectively. The vertical lines in the top panels mark the peak times of the flares. The arrows labeled "1", "2", "3", and "4" in the middle panels denote the erupting materials, the circular ribbons, the parallel ribbons, and the remote brightenings, respectively. The red and white arrows in the bottom panels indicate the bright kernels and large-scale loops, respectively.

**Table 1.** Five homologous M-class flares in AR 11476

| No. | Date | Start time (UT) | Peak time (UT) | End time (UT) | Duration (min) | Class |
| --- | --- | --- | --- | --- | --- | --- |
| Flare 1 | 2012-05-08 | 13:02 | 13:08 | 13:12 | 10 | M1.4 |
| Flare 2 | 2012-05-09 | 12:21 | 12:32 | 12:36 | 15 | M4.7 |
| Flare 3 | 2012-05-09 | 21:01 | 21:05 | 21:09 | 8 | M4.1 |
| Flare 4 | 2012-05-10 | 04:11 | 04:18 | 04:23 | 12 | M5.7 |
| Flare 5 | 2012-05-10 | 20:20 | 20:26 | 20:30 | 10 | M1.7 |

Figure 5 displays the magnetic fields just before the occurrence of each flare. In the regions close to the PILs, the horizontal magnetic fields are strong and highly sheared, as indicated by the lengths and orientations of the blue and red arrows in the top panels. In the squashing factor $Q$ maps (the middle panels) in the photospheric layer, the high-$Q$ regions indicated by green arrows represent the intersections of quasi-separatrix layers (QSL; Demoulin et al. 1996, 1997) with the photosphere. The circular high-$Q$ region denoted by arrow "A" coincides with the circular flare ribbon. The high-$Q$ region denoted by arrow "B" corresponds to the main PIL of the bipole. In the three-dimensional (3D) magnetic fields extrapolated from the NLFFF before each flare, a flux rope lying along the PIL can be identified (see the red curves in the bottom panels). Moreover, there are arch-shaped loops (indicated by the blue curves) overlying the flux ropes.

In the NLFFF model, we search for magnetic null points using the Poincaré index and Newton-Raphson methods (Guo et al. 2017). For Flares 1–4, we find the expected null points in the 3D magnetic fields just before the occurrence of flares. As shown in Figure 6, the positions of the null points are marked with the brown dot symbols. The left and right panels are the top view and side view, respectively. The null points for the four flares are situated at about 3.8 Mm, 5.7 Mm, 3.4 Mm, and 2.6 Mm above the photosphere, respectively. The green curves are the fan-spine skeletons and the red curves represent the flux ropes. The footpoints of the fan structures are just located in the circular-shaped positive magnetic fields surrounding the negative ones. The inclined spine structures extend to the remote negative magnetic fields. The flux ropes are embedded within the fan. This kind of magnetic topology provides the favorable condition for the occurrence of null point reconnection resulting in solar flares. While for Flare 5, we do not find any null point in the pre-flare magnetic fields.

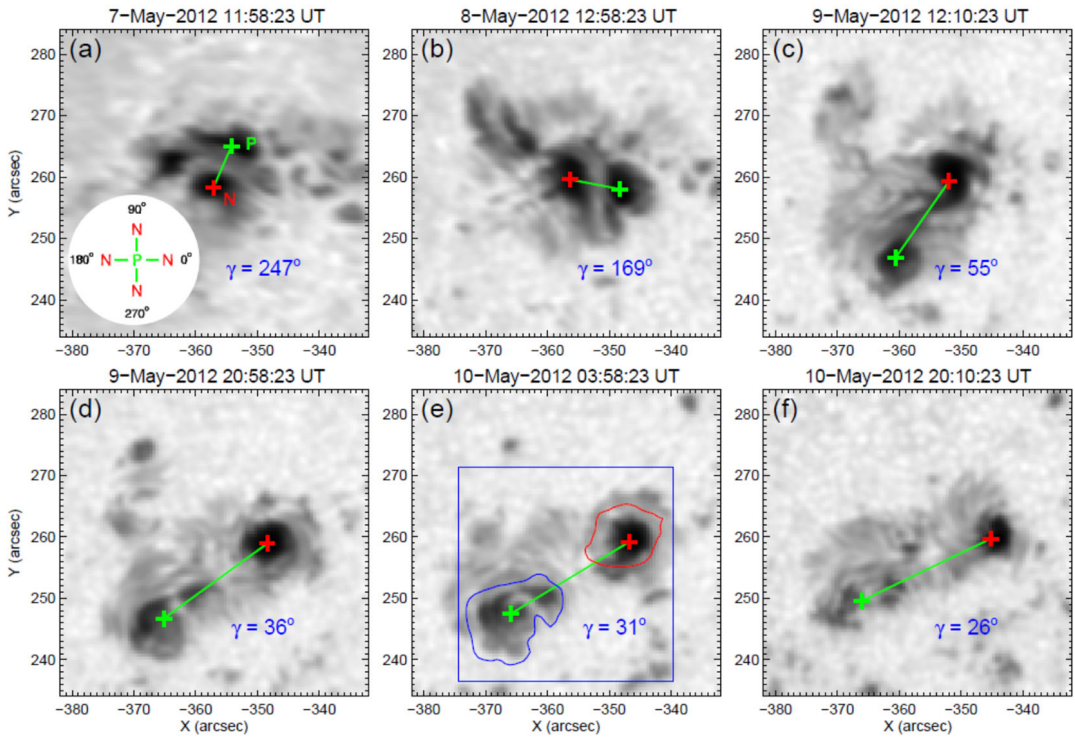

**Fig. 3.** Sequence of HMI intensitygrams displaying the shearing motion of the bipolar sunspots. The green and red plus symbols, connected by the green lines, mark the intensity centroids of the sunspots with positive (P) and negative (N) polarities, respectively. The orientation, $\gamma$, is defined as varying within the range of $[0, 360]°$, as illustrated in the inset circular image in (a). The box in (e) outlines the field of view (FOV) of Figure 4, and the blue and red curves are the contours of the positive and negative magnetic fields at $\pm$ 500 G levels. An animation (Movie1.mp4) of this figure is available.

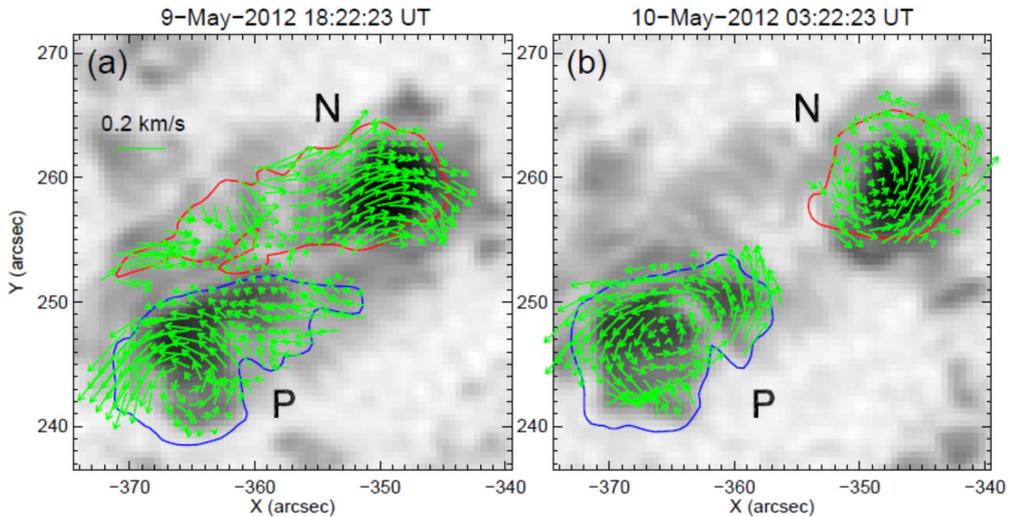

**Fig. 4.** HMI intensitygrams showing the sunspot rotation. The blue and red curves are the contours of the positive and negative magnetic fields at $\pm$ 500 G levels. The arrows indicate the horizontal velocities of the sunspot structures. "P" and "N" indicate the positive and negative polarities, respectively.

In order to further examine the similarity of the five flares, we compare their evolution processes in AIA multi-wavelength observations. For the first four flares with the similar magnetic topologies, their evolution processes were almost identical in AIA UV/EUV images. Figure 7 shows the AIA 1600, 304, and 171 Å images displaying the evolution of Flare 2 as an example (see also Movie 2). At 12:23 UT on May 9, there was a faint brightening (denoted by the arrow in panel (a1)) at the location of a filament (indicated by the arrow in panel (b1)). As the filament went on rise, the brightening was enhanced, and a curved flare ribbon (denoted by the arrow in panel (a2)) was visible at the outskirts. About three and a half minutes later, the curved ribbon extended further to form a circular shape (denoted by the arrow in panel (a3)), and the filament material was erupted outward violently (indicated by the arrow in panel (b3)). Then the brightness at the flare kernel decreased, and a pair of flare ribbons (denoted by the arrows in panels (a4) and (b4)) were prominent at the location previously occupied by the erupting filament. In the 171 Å images, it is clearly shown that the erupting filament (indicated by the green arrow in panel (c4)) was confined by the large-scale loops (denoted by the white arrow in panel (c2)) and moved westward to the remote ends of the large-scale loops.

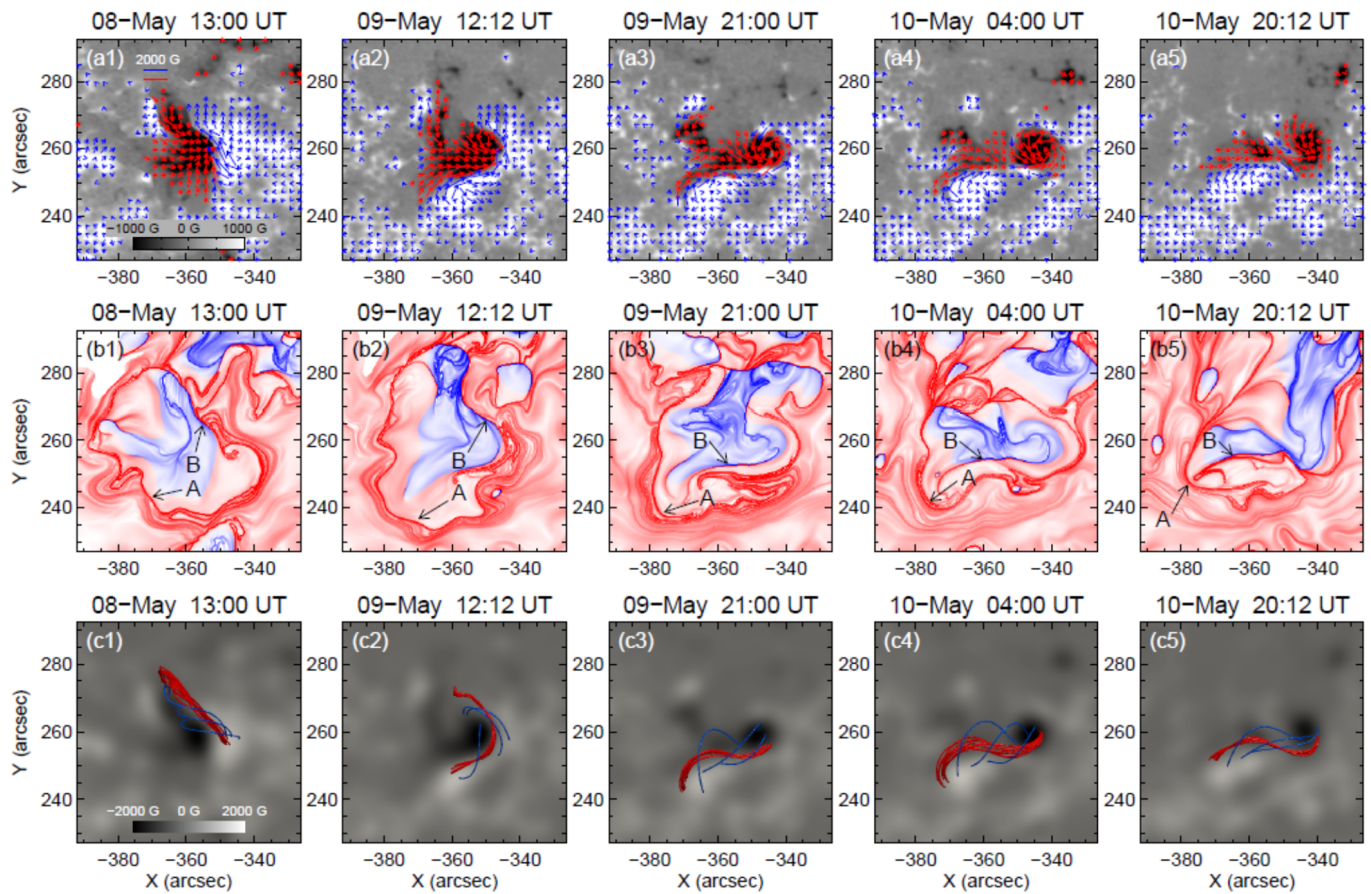

**Fig. 5.** Magnetic fields at the origin site of flares. (a1-a5) HMI vector magnetograms just before the occurrence of five M-class flares. The length and orientation of each blue or red arrow represent the strength and direction of the horizontal magnetic field therein. (b1-b5) Corresponding squashing factor Q maps. The arrows labeled "A" and "B" denote the high Q regions coinciding with the circular ribbon and main PIL of the bipolar region, respectively. (c1-c5) Top view of the reconstructed coronal structures. The red and blue curves represent the magnetic flux ropes and the overlying loops, respectively.

Then one important issue is to compare the fifth flare with the others. As presented in Figure 8 (see also Movie 3), at the beginning of Flare 5, the initial brightening (denoted by the arrow in panel (a1)) also appeared at the site of a filament (denoted by the arrow in panel (b1)). Then a short flare ribbon developed gradually into a circular ribbon (denoted by the red arrow in the top panels). The filament eruption (denoted by the arrow in panel (b3)) resulted in the appearance of a pair of parallel ribbons (indicated by the black arrows) within the circular ribbon. The filament material (denoted by the arrow in panel (c4)) was erupted upward first and then moved toward the west footpoints of the overlying confining loops (indicated by the arrow in panel (c2)). The evolution behavior of the fifth flare was similar to those of the other four flares.

## 4. Conclusions and Discussion

With the multi-wavelength images and vector magnetograms from the SDO, we study the formation process of five M-class circular-ribbon flares which occurred successively at the same location and were not associated with any CME. These five flares had similar properties, forming a series of homologous confined major circular-ribbon flares. Before and during the five flares, the sunspots with opposite polarities sheared against each other and rotated individually. Before each flare, the magnetic fields at the PIL were highly sheared, and a magnetic flux rope lay there with overlying arch-shaped loops. For the first four flares, the magnetic null points were situated at approximately 3.8 Mm, 5.7 Mm, 3.4 Mm, and 2.6 Mm above the photosphere, respectively. However, for the fifth flare, no null point was detected. Nevertheless, in the UV/EUV images, the evolution behaviors of all the flares were almost identical.

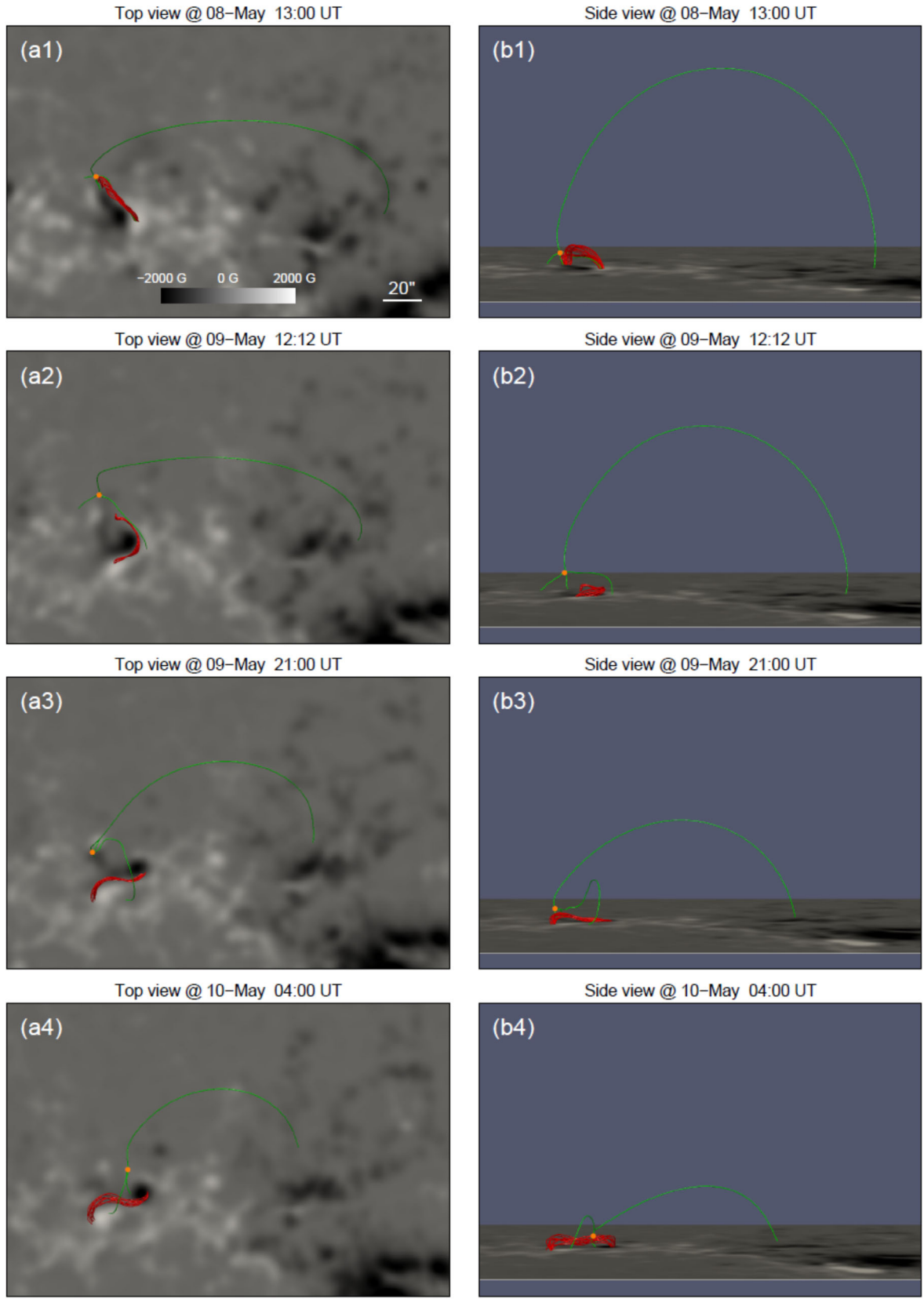

**Fig. 6.** Magnetic topology just before the occurrence of flares. (a1-a4) Top view of the skeletons of the fan-spine represented by green curves together with the null points represented by the brown dot symbols. The red curves represent the magnetic flux ropes within the fan structures, and the backgrounds are vertical magnetograms taken at different times. (b1-b4) Similar to (a1-a4), but from a side perspective.

For these five major flares, some studies have focused on different ones, e.g., Flare 2 (M4.7) was studied by López Fuentes et al. (2018), Flare 3 (M4.1) by Poisson et al. (2020), and Flare

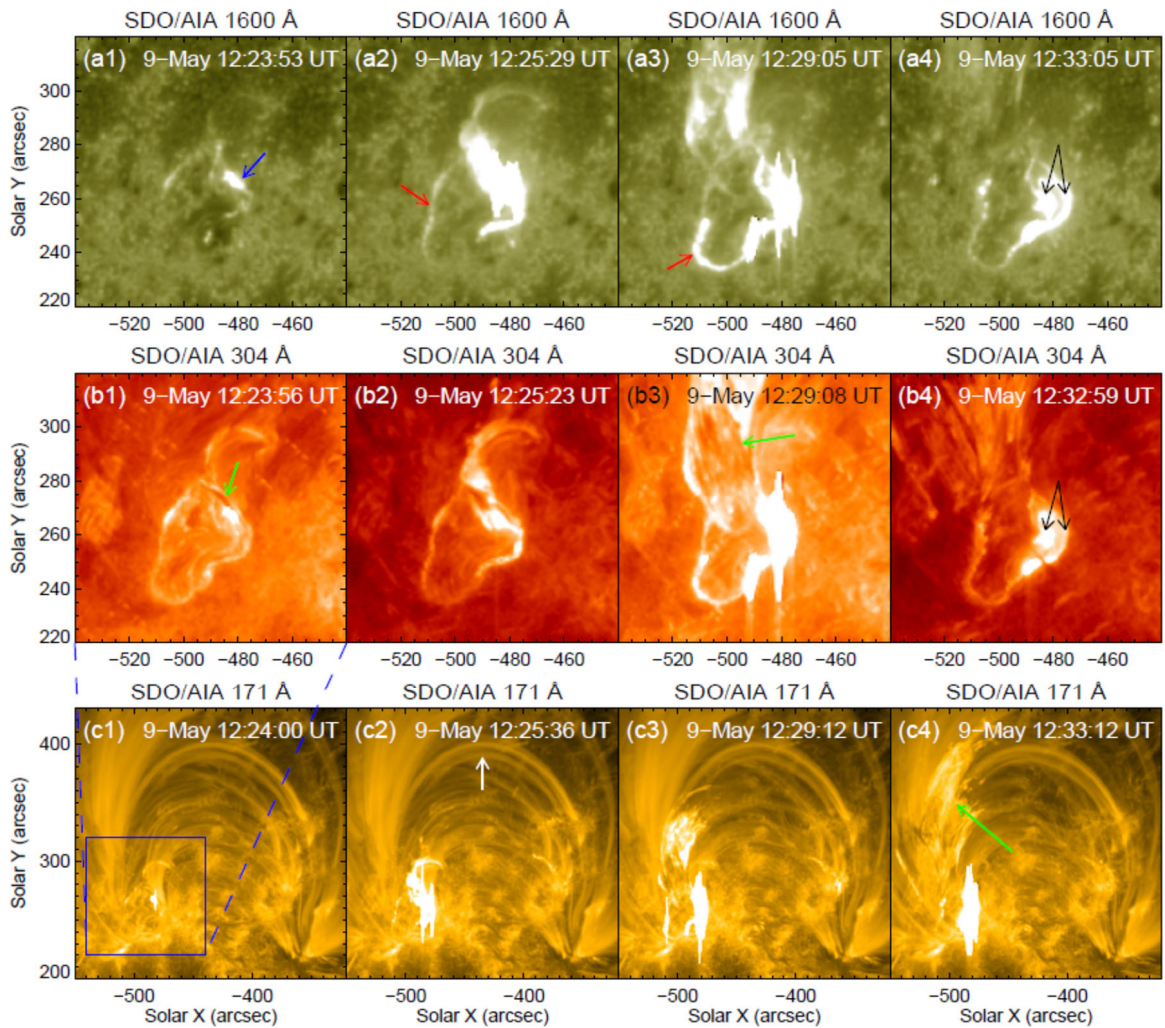

**Fig. 7.** Multi-wavelength images displaying the evolution of Flare 2. (a1-a4) AIA 1600 Å images. (b1-b4) AIA 304 Å images. (c1-c4) AIA 171 Å images. The blue, red, black, green, and white arrows denote the brightening, circular ribbon, a pair of parallel ribbons, filament material, and large-scale loops, respectively. The blue window in (c1) outlines the FOV of the top and middle panels. An animation (Movie2.mp4) of this figure is available.

4 (M5.7) by Yang & Zhang (2018). López Fuentes et al. (2018) described the magnetic field evolution of AR 11476 with a focus on Flare 2. They found that the PIL of the bipole rotated clockwise as the opposite polarities sheared against each other, which refers to the shearing motion between two sunspots. This behavior implies the injection of magnetic helicity, which contributes to the accumulation of free energy (López Fuentes et al. 2024). As demonstrated in the present study, besides the clockwise shearing motion (Figure 3), significant sunspot rotation was observed (Figure 4). The negative polarity sunspot generally rotated counterclockwise, even though it exhibited slight clockwise rotation during some certain periods. While the positive polarity sunspot displayed persistent counterclockwise rotation. Thus the rotational motion of the sunspots is also considered to play an important role in the energy buildup for the flares. This result is consistent with previous studies about the accumulation of flare energy both in observation and simulation (e.g., Yang et al. 2017; Hou et al. 2018; Jiang et al. 2021).

Generally, the formation of a circular-ribbon flare begins with the rising of a flux rope (sometimes appearing as a mini-filament) within the dome-shaped magnetic structure of fan-spine topology (Shen et al. 2019; Yang et al. 2020a). This type of magnetic configuration is favorable for flare occurrence via null point magnetic reconnection (e.g., Antiochos 1998; Aulanier et al. 2000; Cheng et al. 2023). For the flares in this study, we search for null points in the extrapolated magnetic fields. Finally, we find the existence of null points in the coronal magnetic fields before Flares 1–4 (see Figure 6). For Flare 5, no null point is found in the magnetic fields before the flare. The start time of Flare 5 was 20:20 UT on May 10, and the magnetogram used for extrapolation was taken at 20:12 UT. Given the similarity of these flares, we speculate that a null point responsible for the occurrence of Flare 5 may have existed, which might be formed after 20:12 UT and could not be captured due to the 12-min cadence limitation of the SHARP data. For the same AR, López Fuentes et al. (2018) modeled the coronal magnetic fields using the linear force-free field (LFFF) approach with the HMI LOS magnetograms as the bottom boundary. They did not find a null point that could be linked to the flare. This discrepancy may be attributed to the usage of different models and magnetograms. Compared to LFFF extrapolation, NLFFF model better captures the complexity of real coronal magnetic environments by allowing existence of nonlinear concentration of electric currents. However, NLFFF requires computationally intensive iter-ative algorithms, depends on vector magnetogram which has relatively a high noise level, and risks numerical non-uniqueness, whereas LFFF offers simplicity, stability, and rapid solutions for magnetic structure.

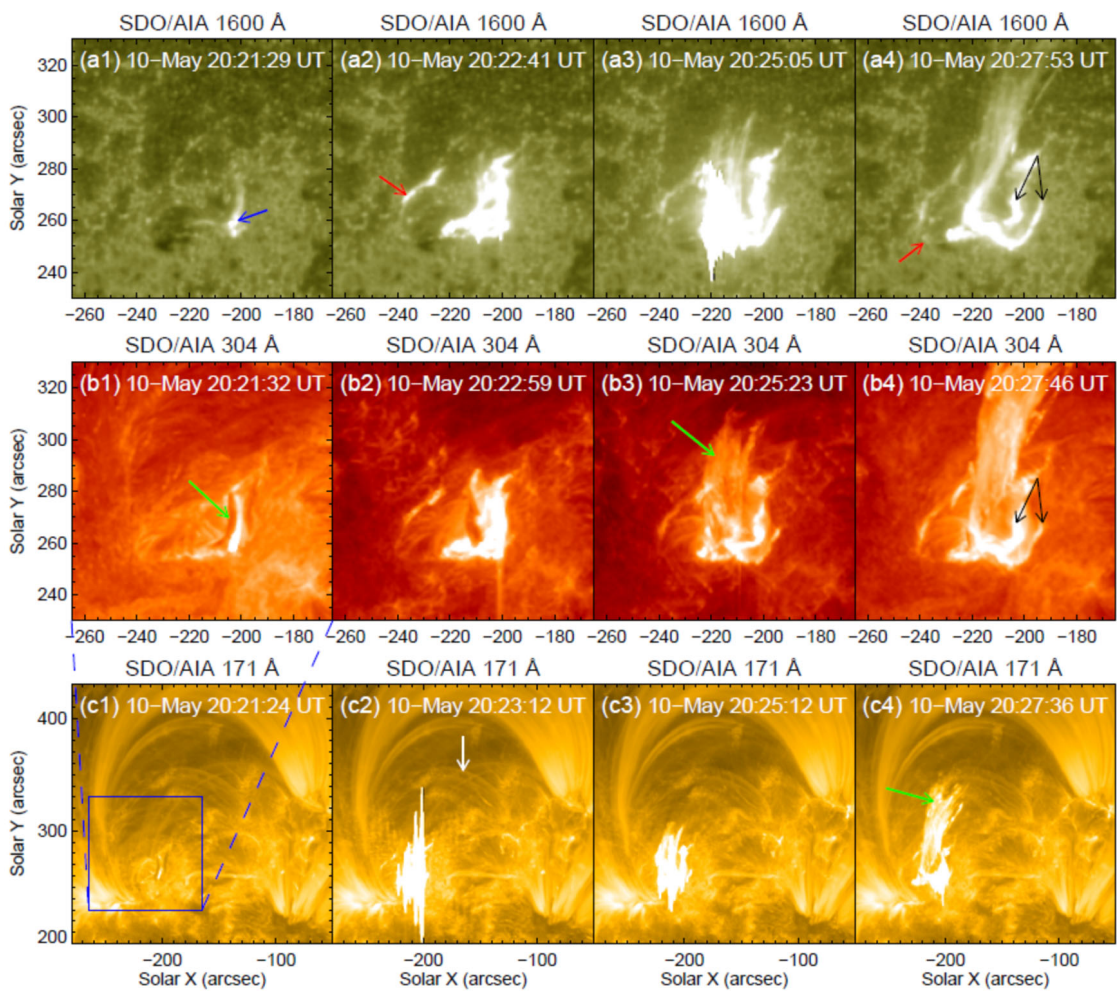

**Fig. 8.** Similar to Figure 7, but for Flare 5. An animation (Movie3.mp4) of this figure is available.

As revealed in the previous studies, if the spine structure of the fan-spine topology is open to the interplanetary space, a jet-like CME may be formed as a result of the filament eruption (Pariat et al. 2009; Shen 2021). However, if the spine extends to a remote region at the solar surface (as shown in Figure 7 of Yang et al. 2020b), the material expelled from the null point may travel to the remote end of the spine causing a remote brightening, rather than being ejected into the interplanetary space. In the present study, all the flares exhibited significant remote brightenings since the erupting filaments were confined by large-scale loops (see Figure 2). Thus, the confined flares were formed, as no CME was associated with them.

In summary, we have analyzed the vector magnetograms with the help of NLFFF model, and provide clear evidence for the drivers of successive, homologous flares as well as the nature of confined events. Our results clarify that, in addition to shearing motion, sunspot rotation also played an important role in energy accumulation for the homologous flares in AR 11476. It is also revealed that the null points of the fan-spine topology were crucial for energy release through magnetic reconnection therein, although no null point was found in some previous work. Large-scale magnetic loops prevented the erupting material from escaping the Sun, thus forming the observed homologous confined major circular-ribbon flares.

*Acknowledgements.* We are grateful to the referee for the valuable comments and suggestions. We thank Profs. Rui Liu and Yang Guo, and Dr. Jun Chen for helpful discussions. This work is supported by the National Key R&D Program of China (2022YFF0503003), Beijing Natural Science Foundation (1252034), the Strategic Priority Program of the Chinese Academy of Sciences (XDB0560000), the National Natural Science Foundations of China (12273060, and 12222306), Shenzhen Key Laboratory Launching Project (ZDSYS20210702140800001), the Specialized Research Fund for State Key Laboratory of Solar Activity and Space Weather, and the Youth Innovation Promotion Association of CAS. The data are used courtesy of the SDO and GOES science teams. SDO is the first mission of NASA's Living With a Star Program.